\PassOptionsToPackage{
  pdfauthor={Inge Madshaven et al.},
  pdftitle={%
	  Photoionization model for streamer propagation mode change
	  in simulation model for streamers in dielectric liquids},
}{hyperref}

\documentclass[10pt]{iopart}
\usepackage{iopams}

\expandafter\let\csname equation*\endcsname\relax
\expandafter\let\csname endequation*\endcsname\relax

\usepackage{etoolbox}                   
\usepackage{amsmath}   
\usepackage[utf8]{inputenc}

\usepackage{lmodern}    
\usepackage{bm}

\usepackage{soul}
\usepackage{color}   

\usepackage{textcomp}     
\usepackage{microtype}

\usepackage{graphicx}
\usepackage{epstopdf}     
\usepackage{dblfloatfix}    
\usepackage{booktabs}       
\usepackage{tabularx}
\usepackage{url} 

\usepackage[squaren]{SIunits}
\usepackage{sistyle}
\newcommand{\si}[1]{\SI{}{#1}}
\renewcommand{\si}[1]{\protect{#1}}

\usepackage[square,numbers,sort&compress]{natbib}

\usepackage[%
  ocgcolorlinks,%
  linktoc=page,%
  breaklinks,%
  hyperfootnotes=false,%
  hidelinks,
  ]{hyperref}

\usepackage{cleveref}
\crefname{equation}{}{}
\Crefname{equation}{Equation}{Equations}
\crefname{figure}{figure}{figures}
\Crefname{figure}{Figure}{Figures}
\crefname{table}{table}{tables}
\Crefname{table}{Table}{Tables}
\crefname{appendix}{}{}
\Crefname{appendix}{}{}

\renewcommand{\vec}[1]{\bm{#1}}
\let\oldhat\hat
\renewcommand{\hat}[1]{\oldhat{\bm{#1}}}

\makeatletter

\begingroup
\catcode`_=\active
\catcode`^=\active
\protected\gdef_{\@ifnextchar'\subtextrm{\@ifnextchar"\subtextsc\sb}}
\protected\gdef^{\@ifnextchar'\suptextrm{\@ifnextchar"\suptextsc\sp}}
\endgroup
\def\subtextrm'#1'{\sb{\textrm{#1}}}
\def\suptextrm'#1'{\sp{\textrm{#1}}}
\def\subtextsc"#1"{\sb{\textsc{#1}}}
\def\suptextsc"#1"{\sp{\textsc{#1}}}
\AtBeginDocument{\catcode`_=12 \mathcode`_="8000}
\AtBeginDocument{\catcode`^=12 \mathcode`^="8000}

\renewcommand{\d}[1]{\text{d}{#1}}

\newcommand{\eez}{\ensuremath{\mathcal{E}_{0}}}                 
\newcommand{\eef}{\ensuremath{\mathcal{E}_{1}}}                 
\newcommand{\een}{\ensuremath{\mathcal{E}_{n}}}                 
\newcommand{\eeg}{\ensuremath{\mathcal{E}_{\gamma}}}            
\newcommand{\eip}{\ensuremath{\mathcal{E}_{\textsc{IP}}}}       %
\newcommand{\eipz}{\ensuremath{\mathcal{E}_{\textsc{IP}}}}      
\newcommand{\efdip}{\ensuremath{\mathcal{E}_{\textsc{fdip}}}}   

\makeatother

\renewcommand{\hl}[1]{#1}
\renewcommand{\colorbox}[2]{#2}

\begin{document}



\title[Photoionization model for streamer propagation mode change]{%
  Photoionization model for streamer propagation mode change
  in simulation model for streamers in dielectric liquids
  }


\author{
  I~Madshaven$^1$,        
  OL~Hestad$^2$,\\        
  M~Unge$^3$,             
  O~Hjortstam$^3$,        
  PO~{\AA}strand$^1$
    \footnote{Corresponding author: \texttt{per-olof.aastrand@ntnu.no}}
  }

\address{%
  $^1$
  Department of Chemistry,
  NTNU -- Norwegian University of Science and Technology,
  7491 Trondheim, Norway
}
\address{%
  $^2$
  SINTEF Energy Research,
  7465 Trondheim, Norway
}
\address{%
  $^3$
  ABB Corporate Research,
  72178 V{\"a}ster{\aa}s, Sweden
}

\begin{abstract}%
%
Radiation is important for the propagation of streamers in dielectric liquids.
Photoionization is a possibility,
but the effect is difficult to differentiate
from other contributions.
In this work,
we model radiation from the streamer head,
causing photoionization when absorbed in the liquid.
We find that photoionization is local in space (\si{\micro m}-scale).
The radiation absorption cross section is modeled
considering that the ionization potential (IP)
is dependent on the electric field.
The result is
a steep increase in the ionization rate when the electric field
reduces the IP below the energy of the first electronically excited state,
which is interpreted as a possible mechanism for changing from slow to fast streamers.
By combining a simulation model for slow streamers based on the avalanche mechanism
with a change to fast mode based on a photoionization threshold for the electric field,
we demonstrate how the conductivity of the streamer channel
can be important for switching between slow and fast streamer propagation modes.
%
\end{abstract}

\vspace{2pc}

\vspace{2pc}
\noindent{\it Date}: \today

%
%
\ioptwocol
%



\section{Introduction}\label{sec:introduction}{

Dielectric liquids 
are widely used in high-voltage equipment,
such as power transformers,
because of their high electrical withstand strength
and ability to act as a coolant%
~\cite{Wedin2014}.
If the electrical withstand strength of the liquid is exceeded,
partial discharges
followed by propagating discharges can occur
and create prebreakdown channels called ``streamers''.
Streamers are commonly classified by
their polarity
and propagation speed,
ranging from
below \SI{0.1}{\kilo\meter\per\second} for the 1st mode
to
above \SI{100}{\kilo\meter\per\second} for the 4th mode%
~\cite{Lesaint2016cxmf}.
Streamers can be photographed by schlieren techniques,
which captures the difference in permittivity
between the gaseous streamer channel and the surrounding liquid~%
\cite{Farazmand1961bhcrhp},
or by capturing light emitted by the streamer%
~\cite{Sakamoto1980dwk579}.
Continuous dim light has been observed from both
the streamer channel and the streamer tip
~\cite{Dung2013czm2},
as well as
bright light from the streamer tip
and re-illuminations of the streamer channel%
~\cite{Linhjell1994chdqcz,Dung2013czm2}.
The intensity of the emitted light
and the occurrence of re-illuminations
increases with higher streamer propagation modes.
Photoionization by light absorbed in the liquid
has been proposed as a possible feed-forward mechanism
involved in the fast 3rd and 4th mode streamers%
~\cite{Linhjell1994chdqcz,Lundgaard1998cn8k5w}.

Streamer propagation is a
multiscale, multiphysics phenomenon
involving numerous mechanisms and processes%
~\cite{Lesaint2016cxmf}.
Developing predictive models and simulations is challenging,
but many attempts exist
{\cite{Biller1996,Beroual2016cxmr}}.
Simulations have often focused on one aspect of the problem,
such as
the electric field~\cite{Niemeyer1984d35qr4,Fofana1998bm4sd5},
production of free electrons~\citep{Kim2008bntb22},
conductance of the streamer channels~\citep{Kupershtokh2006d3d7d3},
inhomogeneities~\citep{Jadidian2014f55gj5},
or
the plasma within the channels~\cite{Naidis2016cxmg}.

In this work we investigate
a model for photoionization%
~\cite{Madshaven2016cx9v,Madshaven2017cx9w}
and combine it with a simulation model
for propagation of streamers through an avalanche mechanism%
~\cite{Madshaven2018cxjf,Madshaven2019c933}.
\hl{%
The simplified cases studied in
~\mbox{\cite{Madshaven2016cx9v,Madshaven2017cx9w}},
mimicking a streamer in a tube,
can give only one streamer mode change.
However,
by not restricting the streamer to a ``tube''
and
including the dynamics of the streamer channel,
we now demonstrate that
the streamer can change between slow and fast mode
multiple times during a simulation.
The present work is organized as follows:
}%
Theory
on
molecular energy states
and
radiation
is given in the next section.
The photoionization model is presented, evaluated and discussed
in sections
\ref{sec:photoionization_model},
\ref{sec:photoionization_results} and
\ref{sec:photoionization_discussion},
respectively.
\Cref{avalanche_model} describes
the simulation model based on electron avalanches,
with photoionization included,
and the results of this model is presented in \cref{avalanche_results}.
The model and the results are discussed in \cref{sec:discussion},
with the main conclusions summarized in \cref{sec:conclusion}.

}


\section{Molecular energy states and radiation}{\label{sec:theory}

Molecules exist in quantum states with different energy $\een$.
Excitation to a state of higher energy
or relaxation to a state of lower energy
can be achieved by
absorbing or emitting a photon,
respectively.
The energy difference between molecular vibrational states
is in the range \si{\milli\electronvolt}
to about \SI{0.5}{\electronvolt},
while molecular electronic states
have energies from some \si{\electronvolt} and up
to around \SI{20}{\electronvolt}.
Change in vibrational states corresponds to infrared (IR) radiation
(room temperature is about \SI{25}{\milli\electronvolt}),
whereas visible (VIS) light (1.7--3.1~\si{\electronvolt})
and ultraviolet (UV) light (above \SI{3.1}{\electronvolt})  
normally correspond to electronic excitations.
The transition probabilities to lower states
gives the lifetime of an excited state,
which varies from \si{\femto\second} to several $\si{\micro\second}$.
In the case of fluorescence,
an excited molecule
relaxes through one or more states,
before relaxing to the electronic ground state.
The final relaxation is the most energetic
and has the longest decay time,
e.g.\
about \SI{7.3}{\electronvolt} and \SI{1}{\nano\second}
in liquid cyclohexane~\cite{Wickramaaratchi1985b724t3}.

The ionization potential (IP) of a molecule
is the energy required to excite an electron
from the ground state \eez\ to
an unbound state.
Applying an external electric field $\vec{E}$
decreases the IP%
~\cite{Smalo2011}
\begin{equation}
    \efdip(E, \theta_'e')
    = \eipz - \beta \cos \theta_'e' \, \sqrt{ \frac{E}{\epsilon_'r' E_{a_0}} }
    \,,
    \label{eq:I_ET}
\end{equation}
where
$\eipz$ is the zero-field IP,
$E_{a_0} = \SI{5.14e11}{\volt\per\meter}$,
$\epsilon_r$ is the relative permittivity of the liquid,
$\cos \theta_'e' = \hat{k}_'e' \cdot \hat{E}$,
and
$\vec{k}_'e'$ is the momentum of emitted electron.
%
\hl{%
The parameter $\beta = {\SI{54.4}{\electronvolt}}$
for the hydrogen atom,
and similar values have been estimated for cyclohexane and
several other molecules%
~\mbox{\cite{Smalo2011}}.
}%
The energy of excited states is usually not significantly affected
by the electric field
in comparison to the field-dependence of the IP%
~\cite{Smalo2011,Davari2013cx9x,Davari2015f6zb32}.

Spectral analysis of the light emitted from streamers
show a broad band of photon energies
up towards 3--4~\si{\electronvolt}%
~\cite{Wong1982df44rx,Bonifaci1991}.
Distinct peaks in the emission spectrum
reveal the presence of entities
such as H$_2$, C$_2$, and CH$_4$,
which are likely products of dissociation and recombination
of hydrocarbon molecules from the base liquid%
~\cite{Wong1982df44rx,Beroual1993bknxqp}.
Stark broadening of the H$_{\alpha}$-line
can be investigated to find
electron densities above
\SI{e24}{\meter^{-3}},
while the relation between
the H$_{\alpha}$ and the H$_{\beta}$-line
point to electron temperatures in the area of \SI{10}{\kilo\kelvin}%
~\cite{Barmann1996fc977m}.
Furthermore,
rotational and vibrational temperatures
of several \si{\kilo\kelvin}
can be estimated from spectral emission
of C$_2$ Swan bands%
~\cite{Ingebrigtsen2008fqzndp}.

During a streamer breakdown,
electrons (and other charged particles)
are gaining energy
and are accelerated
in the electric field.
Energy can be exchanged with other particles through collisions,
possibly resulting in
excitation, ionization or dissociation of molecules.
Subsequently,
relaxation or recombination can cause photon emission.
The radiation $B$
is absorbed by the medium,
given by
$\nabla B = - B \sigma \rho$
\hl{(Beer--Lambert law),}
where
$\sigma$ is the absorption cross section
and
$\rho$ is the number density of the medium.
\hl{%
Integration in spherical symmetry yields
}%
\begin{equation}
    \vec{B}(r)
    = \vec{B_0} \, \left( \frac{r_0}{r}  \right)^2
        \exp \left( - \int_{r_0}^{r} \rho \, \sigma \, \d \ell \right)
        \,,
    \label{eq:sphericalb}
\end{equation}
where
$\vec{B}(r = r_0) = \vec{B_0} = B_0 \hat{r}$.
The ionization cross section of cyclohexane,
for instance,
increases from close to zero below the IP
to about \SI{5e-21}{\meter^2} over the range of around \SI{1}{\electronvolt}%
~\cite{Cool2005fqk5cw}.
For single photons,
cyclohexane begins to absorb around the first excitation energy
and
the absorption cross section increases steadily for higher photon energies%
~\cite{Jung2003fvnn4x}.
A streamer could generate high-energy photons,
which are rapidly absorbed by the liquid
and therefore not measured by experiments%
~\cite{Wong1982df44rx}.

\noindent\colorbox{yellow}{\parbox{\columnwidth}{%
\hspace*{3 ex}
From the radiation $B$,
the photon number density $n_\gamma$ is given by~\cite{Carroll2007}
\begin{equation}
    n_\gamma = B \smash{\big/} \eeg c \,,
    \label{eq:n_gamma}
\end{equation}
%
where
$\eeg$ is the photon energy
and
$c$ is the speed of light in vacuum.
From the Beer--Lambert law and \cref{eq:n_gamma}
it follows that
$\nabla n_\gamma = - n_\gamma \sigma \rho$.
Generally,
$\sigma$ is a superposition of all (absorption) cross sections,
however,
when excitations can be neglected and only ionization is considered,
the ionization rate $W$ (per volume) is given by the change in $n_\gamma$,
\begin{equation}
    W = -\partial_t n_\gamma = n_\gamma \sigma \rho c
    \,,
\end{equation}
where we have used the continuity equation
$\partial_t n_\gamma + \nabla (c n_\gamma) = 0$.
Within a given volume $ \mathcal{V}$,
the rate of ionizing events is $W \mathcal{V}$
and the number of molecules is $\rho  \mathcal{V}$,
which gives the ionization rate per molecule
\begin{equation}
    w(r) = \frac{W \mathcal{V}}{\rho  \mathcal{V}}
         = \int \frac{B(r, \eeg)  \, \sigma(r, \eeg)}{\eeg} \; \d \eeg
    \,,
    \label{eq:w}
\end{equation}
where
we have explicitly stated the radiation and cross section
as functions of the position $r$ and the photon energy $\eeg$.
For instance,
$w = \SI{e-3}{\per\micro\second}$ implies that
$0.1~\%$ of the molecules would be ionized within a \si{\micro\second}.
}}%

}


\section{Defining the streamer radiation model}{\label{sec:photoionization_model}

\begin{figure}[t]
    \centering
    \includegraphics[width=0.85\columnwidth]{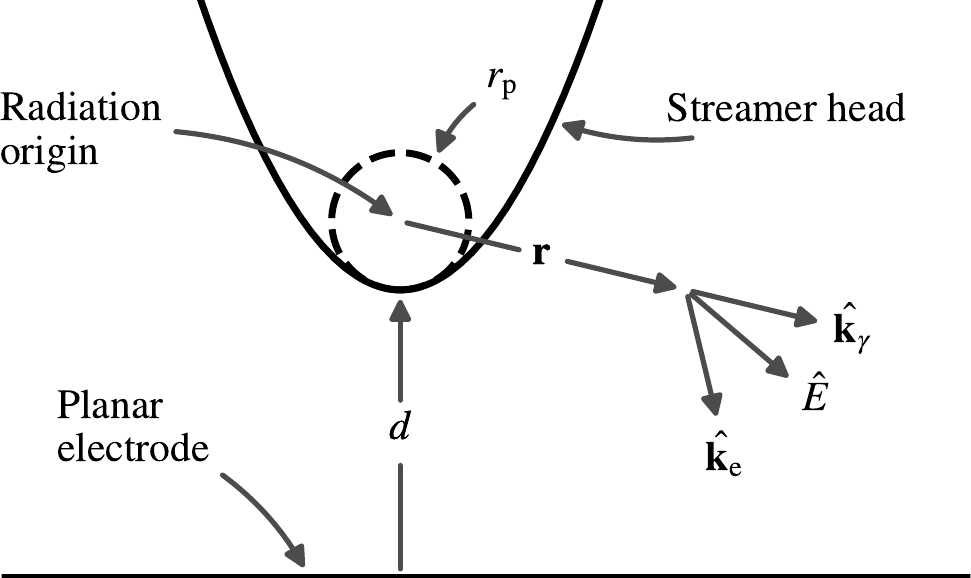}
    \caption{%
    Sketch of a hyperbolic streamer head and relevant variables.
    }
    \label{fig:coordsfinal}
\end{figure}

\begin{figure}[t]
    \centering
    \includegraphics[width=0.97\columnwidth]{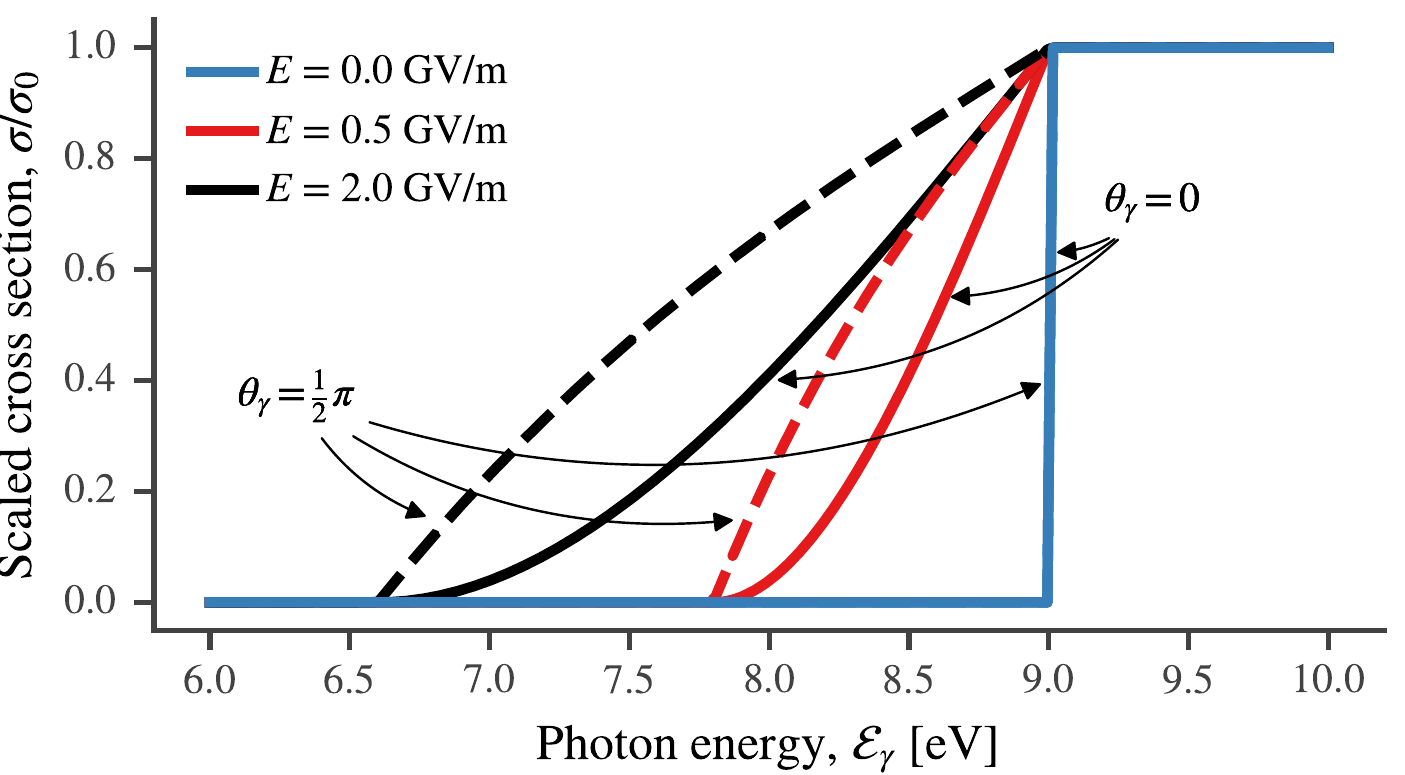}
    \caption{%
        Photoionization cross section $\sigma$
        for different electric fields $E$ and angles $\theta_\gamma$
        as a function of photon energy $\eeg$,
        calculated from
        \cref{eq:sigmaaomega} combined with \cref{eq:I_ET}.
    }
    \label{fig:CSvsF}
\end{figure}

Streamers can emit light sporadically from the channel (re-illuminations)
and continuously from the streamer head,
with fast streamers emitting more light than slow streamers%
~\cite{Linhjell1994chdqcz}.
In this work,
we investigate the possibility of light emitted from the gaseous streamer head
causing ionization in the liquid, resulting in a change to a faster streamer mode.

The probability of emitting the electron in a given direction
is dependent on the momentum of the absorbed photon,
i.e.\ the differential cross section $\d \sigma$
is dependent on the differential solid angle $\d \Omega$,
\begin{equation}
    \d \sigma \propto \sin^2 \theta \, \d \Omega
    \,,
    \label{eq:dsdO}
\end{equation}
where
$\cos \theta = \hat{k}_'e' \cdot \hat{k}_\gamma$.
When $\eeg < \eipz$ we solve for $\eeg = \efdip(E, \Theta)$ in \cref{eq:I_ET}
to find the maximum possible angle $\Theta$ of electron emission.
Then we integrate \cref{eq:dsdO} over all angles
where $\theta < \Theta$ to arrive at
an expression for the photoionization cross section
\begin{align}
    \sigma / \sigma_0
    = 1
    &- \tfrac{1}{4} \big(1 + \cos^2 \theta_\gamma \big)  \big(3\cos \Theta - \cos^3\Theta  \big) \notag\\
    &- \tfrac{1}{2} \sin^2 \theta_\gamma  \cos^3\Theta
    \,,
    \label{eq:sigmaaomega}
\end{align}
where
$\cos \theta_\gamma = \hat{\bm k}_\gamma \cdot \hat{\bm E}$.
\hl{%
Since \mbox{\cref{eq:dsdO}} just gives a proportionality relation,
\mbox{\cref{eq:sigmaaomega}} has been scaled such that
$\sigma (\Theta = \frac{1}{2} \pi, \theta_\gamma) = \sigma_0$.
}%
This is illustrated by \cref{fig:CSvsF},
where
$\sigma = 0$ when $\eeg < \efdip$,
$\sigma = \sigma_0$ when $\eeg > \eipz$,
and dependent on
$\vec{E}$ and $\vec{k}_\gamma$
when $\efdip < \eeg < \eipz$.
\hl{%
For example,
for $\eeg = {\SI{7.5}{eV}}$ and $E = {\SI{2}{GV/m}}$,
we find $\Theta = 0.3 \pi$,
implying that $\efdip(E, \theta_'e' < 0.3 \pi) < \eeg$.
According to \mbox{\cref{eq:dsdO}},
photons with $\theta_\gamma = \frac{1}{2} \pi$ (perpendicular to $\vec{E}$)
have a higher chance of emitting an electron in this region ($\theta_'e' < \Theta$)
than photons with $\theta_\gamma = 0$.
This is reflected in the different cross sections in \mbox{\cref{fig:CSvsF}}.
}

We choose $z = (d + r_'p')$
as the origin of radiation
with a radiance $\vec{B}(r = r_'p') = \vec{B}_0$,
see \cref{fig:coordsfinal}.
Generally,
$B_0$ is comprised of a distribution of photon energies,
however,
we choose to limit the model to only consider
radiation from a single low-energy excited state ($\eeg = \een - \eez$),
since low-energy states are likely the most abundant ones.
Radiation can cause ionization
if the photon energy exceeds the field-dependent IP, i.e.\ $\eeg > \efdip$.
Prolate spheroid coordinates
are used to calculate the Laplacian electric field magnitude and direction%
~\cite{Madshaven2018cxjf},
in order to calculate $\sigma$ by \cref{eq:sigmaaomega}.
The radiance $\vec{B}$ in \cref{eq:sphericalb}
and
the ionization rate $w$ in \cref{eq:w}
can then be calculated,
assuming low density ($\rho \approx 0$) within the streamer head
and constant density in the liquid.
The integration of $\sigma$ is performed numerically
in a straight line from $z = (d + r_'p')$.
Two-photon excitations (absorption to excited states)
and scattering (absorption and re-emission)
are assumed to have low influence and are ignored
in this work.

}


\section{Properties of the radiation model}{\label{sec:photoionization_results}

\begin{figure*}[t]
    \centering
    \includegraphics[width=0.95\linewidth]{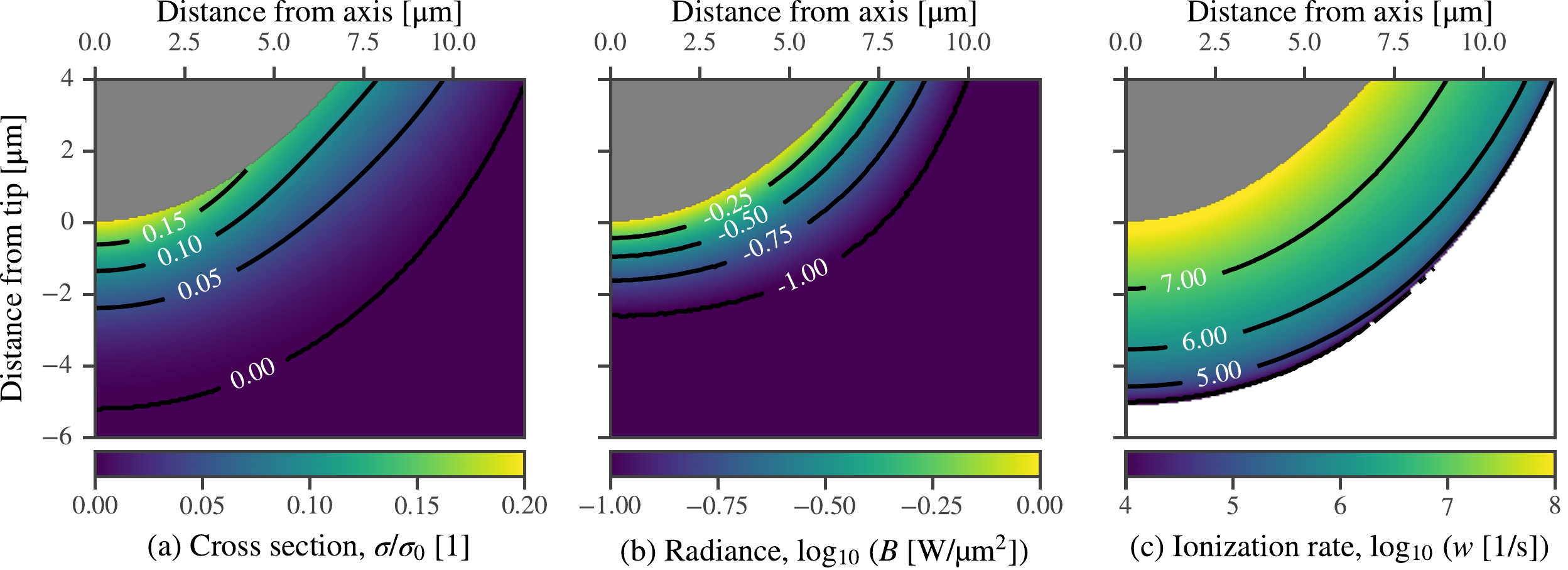}
    \caption{%
        Streamer head (in grey) with
        $r_'p' = \SI{6}{\micro\meter}$
        and
        $V_0 = \SI{100}{\kilo\volt}$,
        placed $d = \SI{10}{\milli\meter}$
        from the planar electrode.
        The cross section (a),
        radiance (b),
        and
        photoionization rate (c)
        are calculated by
        \cref{eq:sigmaaomega},
        \cref{eq:sphericalb},
        and
        \cref{eq:w},
        respectively,
        applying the parameter values
        stated in the first paragraph of \cref{sec:photoionization_results}.
        The $y$-axis is equal for all the plots.
    }
    \label{fig:ceidp_cs_radiance_ionization}
\end{figure*}

\begin{figure*}[t]
    \centering
    \includegraphics[width=1.00\columnwidth]{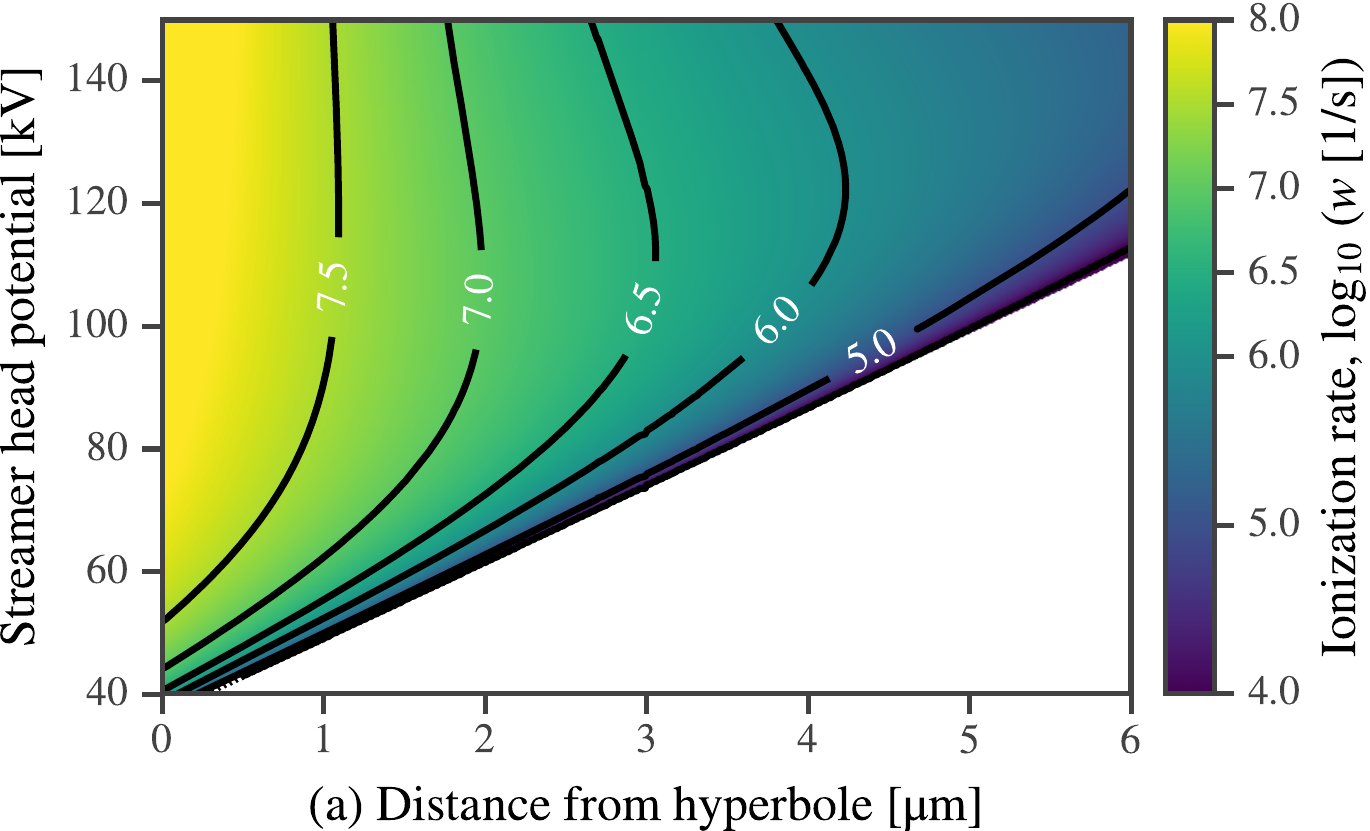}
    \hspace{0.1\columnwidth}
    \includegraphics[width=0.85\columnwidth]{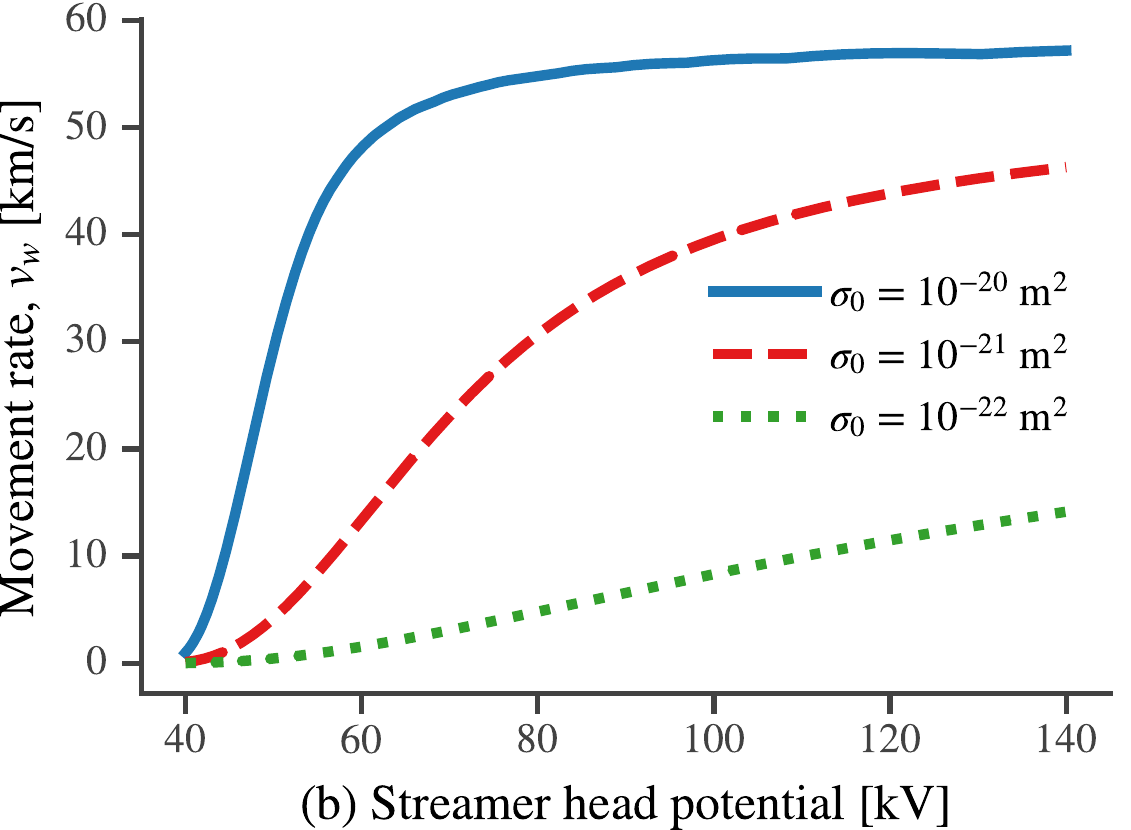}
    \caption{%
        (a)
        Ionization rate $w$ along $z$-axis for different $V_0$.
        \SI{100}{\kilo\volt} corresponds to
        \cref{fig:ceidp_cs_radiance_ionization}(c).
        (b)
        Maximum movement rate $v_w$ calculated by \cref{eq:vw}
        assuming $p = 0.001$.
        The transition is sharpest for $\sigma_0 = \SI{e-20}{m^2}$
        followed by \SI{e-21}{m^2},
        while \SI{e-22}{m^2} resembles a linear increase.
        The magnitude is linearly dependent on $B_0$ and
        inversely dependent on $p$.
    }
    \label{fig:ir_volt_speed}
\end{figure*}

To evaluate the radiation model,
a hyperbolic streamer head
with tip curvature $r_'p' = \SI{6}{\micro\meter}$
is placed with a gap $d = \SI{10}{\milli\meter}$
towards a planar electrode (see \cref{fig:coordsfinal}).
The model liquid is
similar to cyclohexane,
assuming radiation from the lowest excited state,
i.e.\
$\eeg = \eef - \eez = \SI{7}{\electronvolt}$,
$\eipz = \SI{9}{\electronvolt}$,
$\sigma_0 = \SI{e-21}{\meter^2}$
and
$\rho = \SI{5.6e27}{\per\meter^3}$%
~\cite{Madshaven2016cx9v}.
The initial power of the radiation
is set to $B_0 = \SI{1}{W/\micro m^2}$,
which is in the range of the power needed to evaporate the liquid%
~\cite{Gournay1994dw59f5}.
\hl{%
Several other factors,
such as Joule heating,
can contribute to evaporate the liquid,
and the contribution from radiation is unknown.
Furthermore, the
}%
actual radiation power of a streamer is unknown
and likely to fluctuate.
However,
since the results are linear in $B_0$,
setting a value enables
a discussion of whether the results are reasonable.

The area where ionization is possible increases with $V_0$
and covers a range of about
\SI{5}{\micro\meter} from the streamer head
when $V_0 = \SI{100}{\kilo\volt}$,
see \cref{fig:ceidp_cs_radiance_ionization}(a).
At the $z$-axis,
$\sin \theta_\gamma = 0$,
the cross section $\sigma$ is yet the largest
close to the streamer head,
because of the strong electric field $E$.
\Cref{fig:ceidp_cs_radiance_ionization}(a) shows how $\sigma$
declines as the distance from the streamer head increases.
One could expect that $\sigma$ would
decline fast close to the streamer head
as the distance from the $z$-axis increases,
since $E$ declines,
however,
an increase in $\sin \theta_\gamma$
when moving away off-axis
counteracts the reduction in $E$,
resulting in just a slight decrease in $\sigma$.
Numeric integration of $\sigma$ in \cref{fig:ceidp_cs_radiance_ionization}(a)
is applied to find $B$ in \cref{eq:sphericalb},
see \cref{fig:ceidp_cs_radiance_ionization}(b).
The rapid decay of the radiance is expected
considering that $\rho \sigma_0 = \SI{5.6}{\per\micro\meter}$
(i.e.\ a penetration depth of $\delta = 1/\sigma\rho = \SI{0.18}{\micro m}$)
is included in the exponent in \cref{eq:sphericalb}.
The ionization rate per molecule $w$ in \cref{eq:w}
is presented in \cref{fig:ceidp_cs_radiance_ionization}(c).
A major finding is that photoionization is indeed a very local effect
in dielectric liquids,
mainly occurring within a few \si{\micro\meter} of the source,
which is the streamer head in this case.

Increasing $V_0$
increases the ionization rate $w$
close to the streamer head
and increases the reach of the ionization zone
in \cref{fig:ir_volt_speed}(a)
\hl{%
up to about {\SI{100}{kV}}.
At higher potentials,
the ionization rate at a distance of some {\si{\micro m}} decreases
since much of the radiation is absorbed within the first {\si{\micro m}}.
This is evident from the contour for $w = \SI{e6}{/s}$, for instance.
}%
%
%
We may hypothesize that photoionization cause streamer propagation
once a degree of ionization $p$ is obtained.
\hl{%
The time $t_w$ required to reach $p$ is
$t_w = p/w$,
and this time varies with the distance from the streamer head $\Delta r$.
Both the time and the distance are important,
obtaining $p$ fast very close to the streamer
have to be weighed against having a longer $t_w$
at a distance further from the streamer.
As such, we define the photoionization speed of the streamer,
}%
\colorbox{yellow}{\parbox{\columnwidth}{%
\begin{equation}
    v_w = \textrm{max} \left\{\frac{\Delta r}{t_w} \right\}
        = \textrm{max} \left\{\frac{\Delta r w}{p} \right\}
    \,.
    \label{eq:vw}
\end{equation}
}}
\hl{%
The speed $v_w$ is set to the maximum value of the product
of $\Delta r$ and $w$,
where $w = w(\Delta r)$ is calculated numerically,
for a range of $\Delta r$ close to the streamer head.
}%
Since measured electron densities in streamers
point to a degree of ionization in the range of 0.1~\% to 1~\%%
~\cite{Barmann1996fc977m,Ingebrigtsen2008fqzndp},
we assume that $p = 0.001$ is required for propagation.
The photoionization speed $v_w$ of the data
in \cref{fig:ir_volt_speed}(a)
is presented in \cref{fig:ir_volt_speed}(b),
showing an increase in $v_w$
as $\efdip$ is reduced below $\eeg$.
\hl{%
Changing to $p = 0.01$ would yield the same result
if also $B$ was increased tenfold
since the magnitude of $v_w$ is dependent on the radiated power
($v_w \propto w \propto B_0$).
Neither the value of $B_0$ or $p$ is known and
we cannot assert that photoionization indeed leads
to such a drastic speed increase $v_w$ as shown in \mbox{\cref{fig:ir_volt_speed}(b)},
however,
the important part of the model is to show
that photoionization can be affected by the electric field strength
and
that the effect is local.
}%
Physically,
when the liquid no longer can absorb light to a bound excited state,
the result is direct ionization,
and
it is reasonable that ionization contributes more to the propagation speed
than emission of light or local heating.
The transition from low to high speed
(low to high ionization rate)
in \cref{fig:ir_volt_speed}(b)
for the largest cross section ($\sigma_0 = \SI{e-20}{m^2}$)
occurs over a short voltage range of about \SI{20}{kV}.

}


\section{Discussion of the radiation model}{\label{sec:photoionization_discussion}

The modeled photoionization cross section increased
from zero towards a maximum of \SI{e-21}{m^2},
which resulted in rapid absorption within a few \si{\micro m}.
The real absorption might be even more rapid,
since the cross section of cyclohexane is about 5 times larger
for ionizing radiation~\cite{Cool2005fqk5cw}.
\hl{%
An increase in the cross section to $\sigma_0 = {\SI{5e-21}{m^2}}$
gives a shorter penetration depth $\delta = 1/\sigma\rho$,
which results in even shorter range for the radiance and ionization
than shown in \mbox{\cref{fig:ceidp_cs_radiance_ionization}}.
According to \mbox{\cref{fig:ir_volt_speed}(b)}
this gives a steep increase in the movement rate $v_w$.
}%
The fluorescence of cyclohexane is
consistent with radiation from the first excited state~\cite{Wickramaaratchi1985b724t3},
but the absorption to this state is intrinsically low~\cite{Jung2003fvnn4x}.
Radiation from fluorescence may thus transport energy
away from the streamer head.

Excited molecules in the liquid have a high probability of non-radiative
relaxation which heats the liquid.
In strong electric fields,
the IP is reduced
and bound excited states become unbound,
i.e.\ they appear above the ionization threshold~\cite{Davari2013cx9x},
and instead of heating,
absorption causes ionization.
It is,
however,
difficult to assess how
an electric field affects cross sections.
By assuming an increase in the cross section when the field is increased
(see \cref{fig:CSvsF}),
more radiation is absorbed,
but the effect also becomes more local.
The model therefore predicts
a faster propagation when the radiation from the streamer head
is absorbed directly in front of the streamer,
in line with \cref{fig:ir_volt_speed}(b)
where higher cross sections results in higher speeds.

The photoionization cross sections $\sigma$
for linear alkanes and aromatics
differ by more than a decade,
from about \SI{1e-22}{m^2} to \SI{5e-21}{m^2}%
~\cite{Eschner2011dfh4qv}.
Given a number density $\rho = \SI{5e27}{/m^3}$,
the penetration depth $\delta$
is between \SI{2}{\micro m} and \SI{0.04}{\micro m}, respectively.
Ionizing radiation emitted when electrons recombine with cations
is therefore rapidly absorbed,
however,
non-ionizing radiation having lower absorption cross section can propagate further.
If we assume that fluorescent radiation from cyclohexane
is absorbed with a cross section of 1/100 of the ionizing radiation,
this radiation has a reach of several \si{\micro m}.
In combination with a low-IP aromatic additive,
having a larger cross section,
the reach of the radiation is reduced,
but radiation absorbed by the additive causes ionization
whereas absorption to cyclohexane results in heat.
For instance,
pyrene ($\eip = \SI{7}{eV}$~\cite{Davari2015f6zb32}) is ionized when
absorbing fluorescent radiation from cyclohexane,
and could facilitate streamer growth
by providing seed electrons for new avalanches.
A similar result is found for gases
where additives absorbing ionizing radiation
can increase the streamer propagation speed
for a single branch%
~\cite{Naidis2018dbjz}.
\hl{%
Furthermore,
excited states of the additives can have lifetimes of tens to hundreds of ns%
~\mbox{\cite{Brownrigg2009bk5fkn}},
which increases the probability of two-photon ionization
compared with a lifetimes up to about a ns in pure cyclohexane
~\mbox{\cite{Wickramaaratchi1985b724t3}}.
}%
\hl{%
As such, low-IP additives can facilitate slow streamers by
reducing the inception voltage,
increase the propagation length,
and
reduce the breakdown voltage~\mbox{\cite{Lesaint2000c4xf84}}.
Facilitated growth can lead to more branching,
which is possibly why such additives can increase the acceleration voltage~%
\mbox{\cite{Lesaint2000c4xf84}}.
Increased branching can stabilize the streamer through electrostatic shielding,
however,
photoionization in front of the streamer can be involved in a change to a fast mode~%
\mbox{\cite{Linhjell1994chdqcz,Lundgaard1998cn8k5w}}.
For instance,
if one branch escapes the shielding from the others,
the electric field surrounding it would increase,
reducing the IP and allowing more of the radiation to cause ionization.
}%

\hl{%
Under normal conditions,
electrical insulation in liquids is
a steady-state process where the added energy
by the applied electrostatic potential is released
through radiation as either heat or light in the UV/VIS region.
Similarly,
during a streamer breakdown,
the added energy can dissipate in the liquid,
but also cause streamer propagation
when the energy dissipation is concentrated.
The availability of electronic excited states is therefore crucial,
and because of the strong field-dependence of the IP,
the number of available excited states decrease with increasing field~%
\mbox{\cite{Davari2013cx9x}}.
Additives with lower excitation energies,
sustaining to higher fields,
may therefore be an approach to increase the acceleration voltage,
as indicated experimentally~\mbox{\cite{Unge2013c9rd,Liang2019c6v7}}.
}%
\hl{%
The available excited states and absorption probabilities
are therefore important to consider.
One additive that has been studied~\mbox{\cite{Lesaint2000c4xf84}},
pyrene,
has excited states between 4 and {\SI{7}{eV}} (in gas)~%
\mbox{\cite{Davari2015f6zb32}}
}%
and can thus absorb and radiate energy
which is generally not absorbed by cyclohexane.
Pyrene and dimethylaniline (DMA) have a similar $\eip$ and first excitation energy,
and both additives
increase the acceleration voltage in cyclohexane~\cite{Lesaint2000c4xf84,Linhjell2011bjpvr2}.
However,
whereas pyrene absorbs radiation at the lowest excitation energy
which is a $\pi$ to $\pi^*$ transition,
the lowest excited state of DMA is non-absorbing~%
\cite{Galvan2009djk9xp}
and thus the second lowest excitation energy should be considered instead.
\hl{%
This increases the excitation energy from 4 to {\SI{5}{eV}}~%
\mbox{\cite{Galvan2009djk9xp}}.
}%
It is not uncommon that the lowest state is non-absorbing.
For example in azobenzenes,
also studied as an additive in streamer experiments~%
\cite{Unge2013c9rd},
the lowest
$n$ to $\pi^*$ transition is non-absorbing,
whereas the second excitation,
$\pi$ to $\pi^*$,
has a high absorbance and gives the molecules their color%
~\cite{Astrand2000bm5z9v}.
\hl{%
Excited states most likely play a role both in
collision events with primary electrons (affecting impact ionization)
and in absorption of light (affecting photoionization),
but the different contributions are difficult to disentangle
from other mechanisms.
In the end,
which effects that are significant under realistic conditions
need to be established by cleverly designed experiments.
}%

There is a relatively small number of electronic states available below the IP,
but a large number of states above the IP,
often considered as a continuum.
This makes the cross section for ionization
larger than the cross section for absorption
to a bound excited state.
Consequently,
as the IP decreases with an increase in the electric field,
the cross section at certain energies increases.
A local electric field in excess of \SI{0.5}{GV/m}
is sufficient to remove all excited states of cyclohexane in gas phase~\cite{Davari2013cx9x}.
In a liquid where $\epsilon_r = 2$,
we find that a local field of \SI{1.4}{GV/m} reduce the IP by \SI{2}{eV}
from \cref{eq:I_ET},
%
%
which is sufficient to reduce $\efdip$ below the first excited state in cyclohexane.
When the electric field is above this threshold,
cyclohexane cannot absorb radiation to a bound state and is ionized instead.
For a hyperbolic streamer head with $r_'p' = \SI{6}{\micro m}$
in a gap $d = \SI{10}{mm}$,
this threshold is reached at a potential of \SI{37}{kV},
assuming that the local field is the same as the macroscopic field,
and the transition in speed occurs above this in \cref{fig:ir_volt_speed}(b).
The threshold is close to the acceleration voltage in a tube~\cite{Linhjell2013cx9t},
but much lower than the acceleration voltage in a non-constricted large gap~\cite{Lesaint2000c4xf84}.
However, the actual tip radius of the streamer and the degree of branching
are important when calculating the tip field,
as well as space charge generated in the liquid.
Furthermore,
the local field can differ from the macroscopic field.
For instance,
the field is increased by a factor of 1.3
in a spherical cavity in a non-polar liquid~\cite{Smalo2011}.
The model mainly demonstrates how rapid ionizing radiation
(high cross section) is absorbed in the liquid.

}


\section{Avalanche model with photoionization}{\label{avalanche_model}

In earlier work we have developed a model
for simulating the propagation of positive streamers
in non-polar liquids
through an electron avalanche mechanism%
~\cite{Madshaven2018cxjf,Madshaven2019c933}.
Here we incorporate the photo\-ionization mechanism into the streamer model.
A short overview of the model is given below.

Simulation parameters are similar with those used in our previous works~%
\cite{Madshaven2018cxjf,Madshaven2019c933},
i.e.\ a needle-plane gap with cyclohexane as a model liquid.
The needle is represented by a hyperbole (see \cref{fig:coordsfinal})
with tip curvature $r_'n' = \SI{6.0}{\micro\metre}$,
placed $d = \SI{10}{\milli\metre}$
above a grounded plane.
The potential $V_0$ applied to the needle
gives rise to an electric field $\vec{E}$ in the gap.
The Laplacian electric field
is calculated analytically in prolate spheroid coordinates.
Electrons detach from anions in the liquid
(assumed ion density $n_'ion' = \SI{2e12}{\metre^{-3}}$)
and grow into electron avalanches
if the field is sufficiently strong.
The number of electrons $N_'e'$ in an avalanche is given by
\begin{equation}
    \ln N_'e' = \sum_i E_i \, \mu_'e' \, \alpha_'m' \, e^{-{E_\alpha} / {E_i}} \Delta t \,
    \,,
\end{equation}
where
$\alpha_'m' = \SI{130}{/\micro m}$
and
$E_\alpha = \SI{1.9}{GV/m}$ for cyclohexane~\cite{Naidis2015gbf7x2},
$\mu_'e' = \SI{45}{\milli\metre^{2}\per{\volt\second}}$ is the electron mobility,
$i$ denotes a simulation iteration,
and $\Delta t = \SI{1}{ps}$ is the time step.
If an avalanche obtains a number of electrons $N_'e' > 10^{10}$,
it is considered ``critical''.
The streamer grows by placing
a new streamer head wherever an avalanche becomes critical.
Each streamer head, an extremity of the streamer,
is represented by a hyperbole
with tip curvature $r_'s' = \SI{6.0}{\micro\metre}$.
After the inception of the streamer,
the electric potential $V$
and the electric field $\vec{E}$
for a given position $\vec{r}$
is calculated by a superposition
of the needle and all the streamer heads,
\begin{equation}
    V(\vec{r}) = \sum_i k_i V_i(\vec{r})
    \,,
    \quad
    \vec{E}(\vec{r}) = \sum_i k_i \vec{E}_i(\vec{r})
    \,,
\end{equation}
where $i$ denotes a streamer head.
The coefficients $k_i$ correct
for electrostatic shielding between the heads.
Whenever a new head is added,
the streamer structure is optimized,
possibly removing one or more existing heads.
Streamer heads within
$\SI{50}{\micro\metre}$
of another head closer to the plane,
and heads with $k_i < 0.1$,
are removed%
~\cite{Madshaven2018cxjf}.

Each streamer head is associated
with a resistance in the channel towards the needle
and a capacitance in the gap towards the planar electrode%
~\cite{Madshaven2019c933}.
The resistance $R$ and capacitance $C$ is given by
\begin{equation}
    R \propto \ell
    \,,\;\text{and} \;
    C \propto \left( \ln \frac{4z + 2 r_'s'}{r_'s'} \right)^{-1}
    \,,
\end{equation}
where
$\ell$ is the distance from the needle to the streamer head
and
$z$ is the position of the streamer head in the gap.
New streamer heads are given a potential
which magnitude depends on their position
as well as the configuration of the streamer.
The potential $V_i$ of each streamer head is relaxed
towards the potential of the needle electrode $V_0$
each simulation time step.
This is achieved by reducing the difference in potential,
\begin{equation}
    \Delta V_i = V_0 - V_i
    \;\rightarrow\;
    V_i = V_0 - \Delta V_i e^{- \Delta t / \tau_i}
    \,,
\end{equation}
where
the time constant is given by $\tau = \tau_0 RC$
and
$\tau_0 = \SI{1}{\micro s}$.
If the electric field within the streamer channel
$E_'s' = \Delta V_i \smash{\big/} \ell_i$
exceeds a threshold $E_'bd'$,
a breakdown in the channel occurs,
equalizing the potential of the streamer head and the needle.
A channel breakdown affects the potential of a single streamer head
since each streamer head is ``individually'' connected to the needle%
~\cite{Madshaven2019c933}.

Calculating the photoionization cross section in \cref{eq:sigmaaomega}
is a computational expensive operation,
contrary to our avalanche simulation model
which is intended to be
relatively simple and computationally efficient.
The photoionization model indicates
an increase in speed (see \cref{fig:ir_volt_speed})
when $\efdip < \een$ over a short distance into the liquid.
To model photoionization in an efficient way,
we add a ``photoionization speed'' $v_w$
to each streamer head exceeding a threshold $E_w = \SI{3.1}{GV/m}$.
This is implemented by moving such streamer heads
a distance $\vec{s}_w = v_w \Delta t \, \hat{z}$.
\Cref{eq:vw} predicts a speed $v_w$ given a set of parameter values
(see \cref{fig:ir_volt_speed}(b)),
where some,
such as radiation power and degree of ionization,
are unknown.
The chosen power of \SI{1}{W/\micro m^2}
exceeds \SI{100}{W} in total when
distributed over a streamer head with a radius of some \si{\micro m}.
Since a streamer requires about \SI{5}{mJ/m}
for propagation~\cite{Gournay1994dw59f5},
the expected speed exceeds \SI{20}{km/s},
which is in line with \cref{fig:ir_volt_speed}(b).
We choose $v_w = \SI{20}{km/s}$ for the simulations,
which is the order of magnitude given by \cref{fig:ir_volt_speed}(b),
but slow compared to some 4th mode streamers exceeding \SI{100}{km/s}.
However,
this is sufficient
to investigate transitions between slow and fast mode
since it is more than an order of magnitude
above the speed predicted by the simulations
without a photoionization contribution%
~\cite{Madshaven2018cxjf}.

}


\section{Results from avalanche model with photoionization}{\label{avalanche_results}

\begin{figure}
    \centering
    \includegraphics[width=0.97\linewidth]{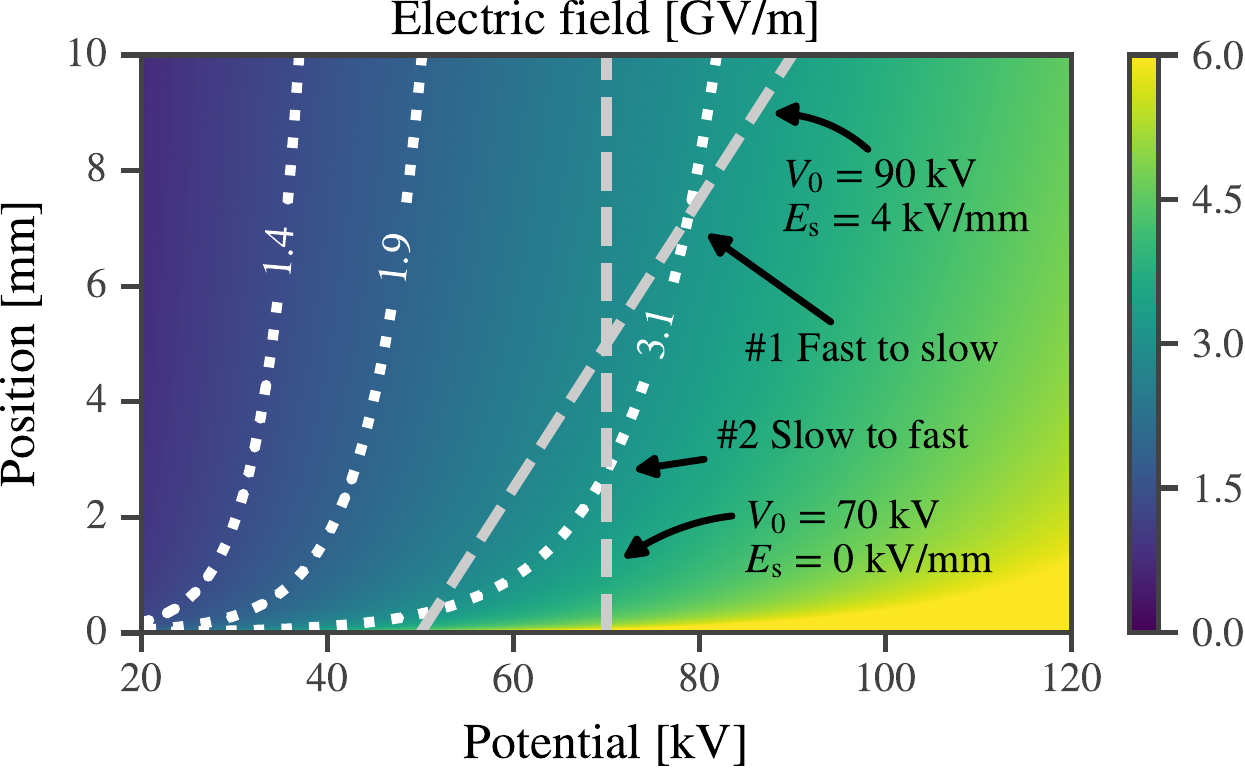}
    \caption{%
        Electric field strength
        at the tip of an electric hyperboloid
        with a tip curvature of \SI{6.0}{\micro\metre}
        for a given position and potential.
        The dotted white lines show the electric field thresholds
        for IP reduction by
        \SI{2}{eV} (\SI{1.4}{GV/m})
        and
        \SI{3}{eV} ($E_w = \SI{3.1}{GV/m}$),
        as well as
        $E_\alpha = \SI{1.9}{GV/m}$.
        The dashed gray lines represent streamers.
        (1) indicate how
        an electric field of $E_s = \SI{4}{kV/mm}$
        can change the propagation mode from fast to slow
        in the beginning of the gap.
        (2) indicate how a highly conducting streamer
        can change from slow to fast towards the end of the gap.
        }
    \label{fig:move_photo_V120kV_d10mm}
\end{figure}

\begin{figure*}
    \centering
    \includegraphics[width=0.97\columnwidth]{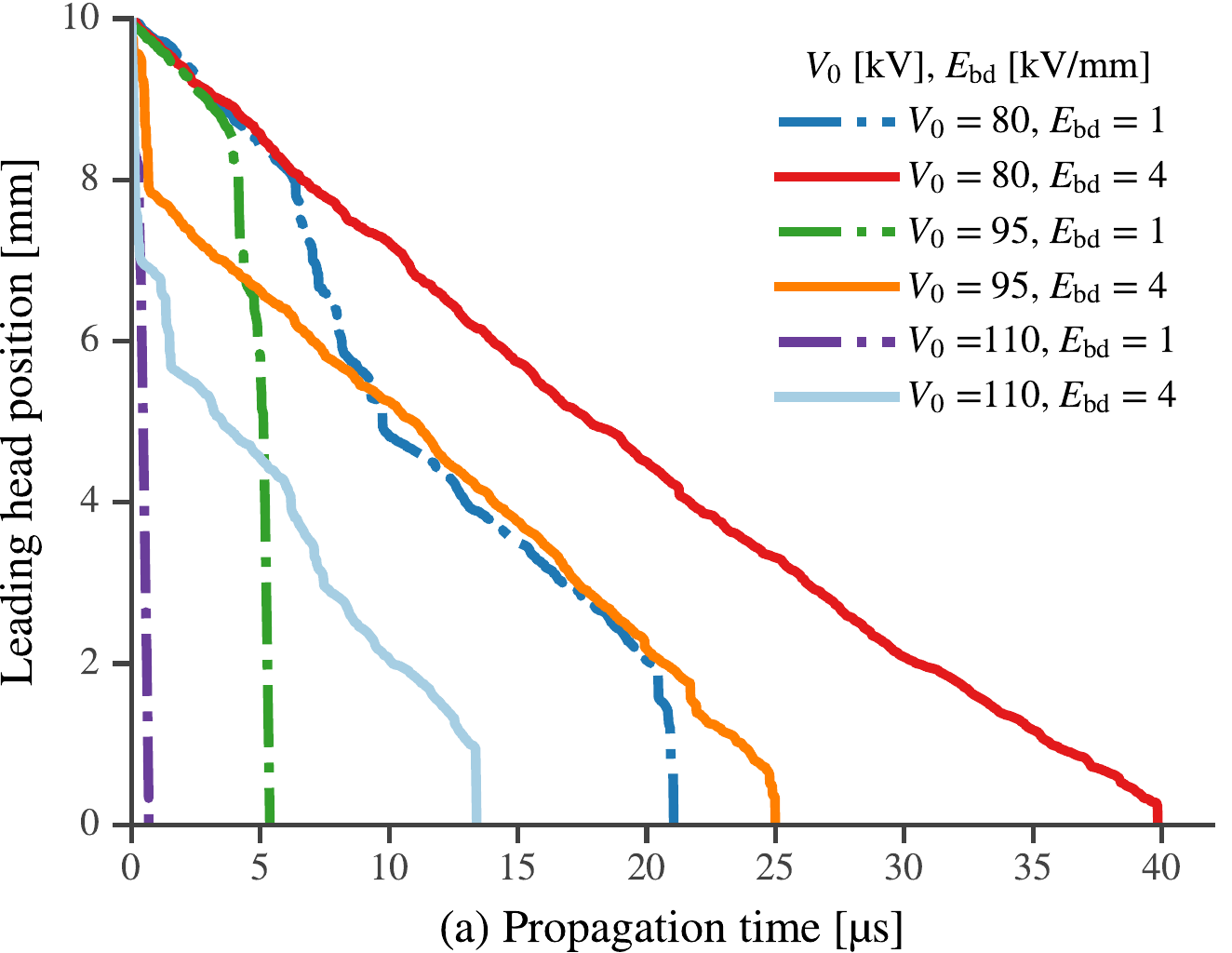}
    \hfill
    \includegraphics[width=0.97\columnwidth]{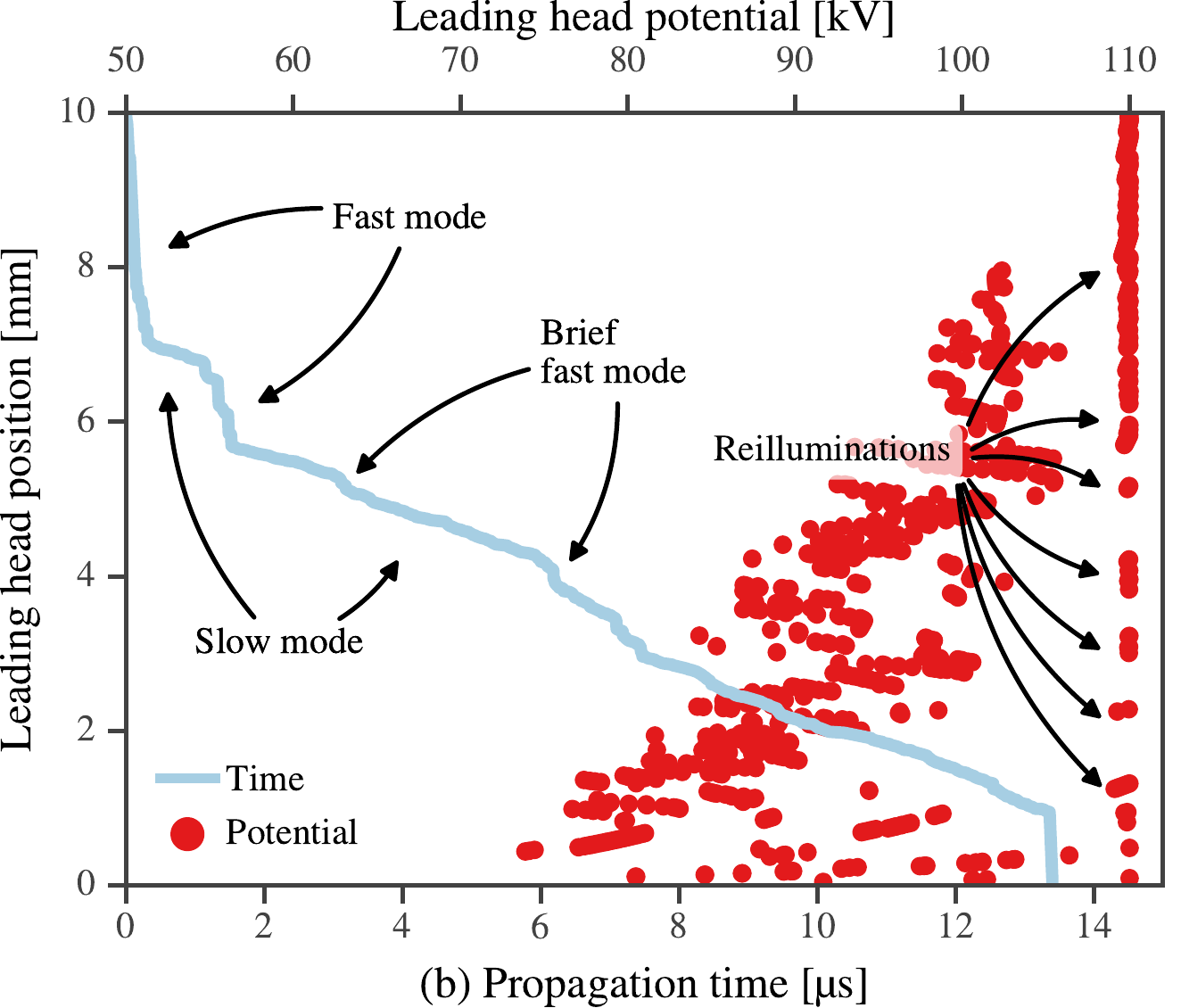}
    \caption{%
        (a)
        Streak plots showing the position of the leading streamer head as a function of time.
        The transition from slow to fast or from fast to slow
        is dependent on the needle potential and the channel breakdown threshold.
        (b)
        Propagation time and leading head potential,
        \hl{each}
        as a function of leading head position in the gap,
        for $V_0 = \SI{110}{kV}$ and $E_'bd' = \SI{4}{kV/mm}$ in (a).
        \hl{%
        The above data is generated by sampling
        the position and potential of the leading streamer head (closest to the opposing electrode)
        for every {\SI{10}{\micro m}} of propagation of each simulation.
        }%
        }
    \label{fig:streaks_potential}
\end{figure*}

\begin{figure}
    \centering
    \includegraphics[width=0.97\linewidth]{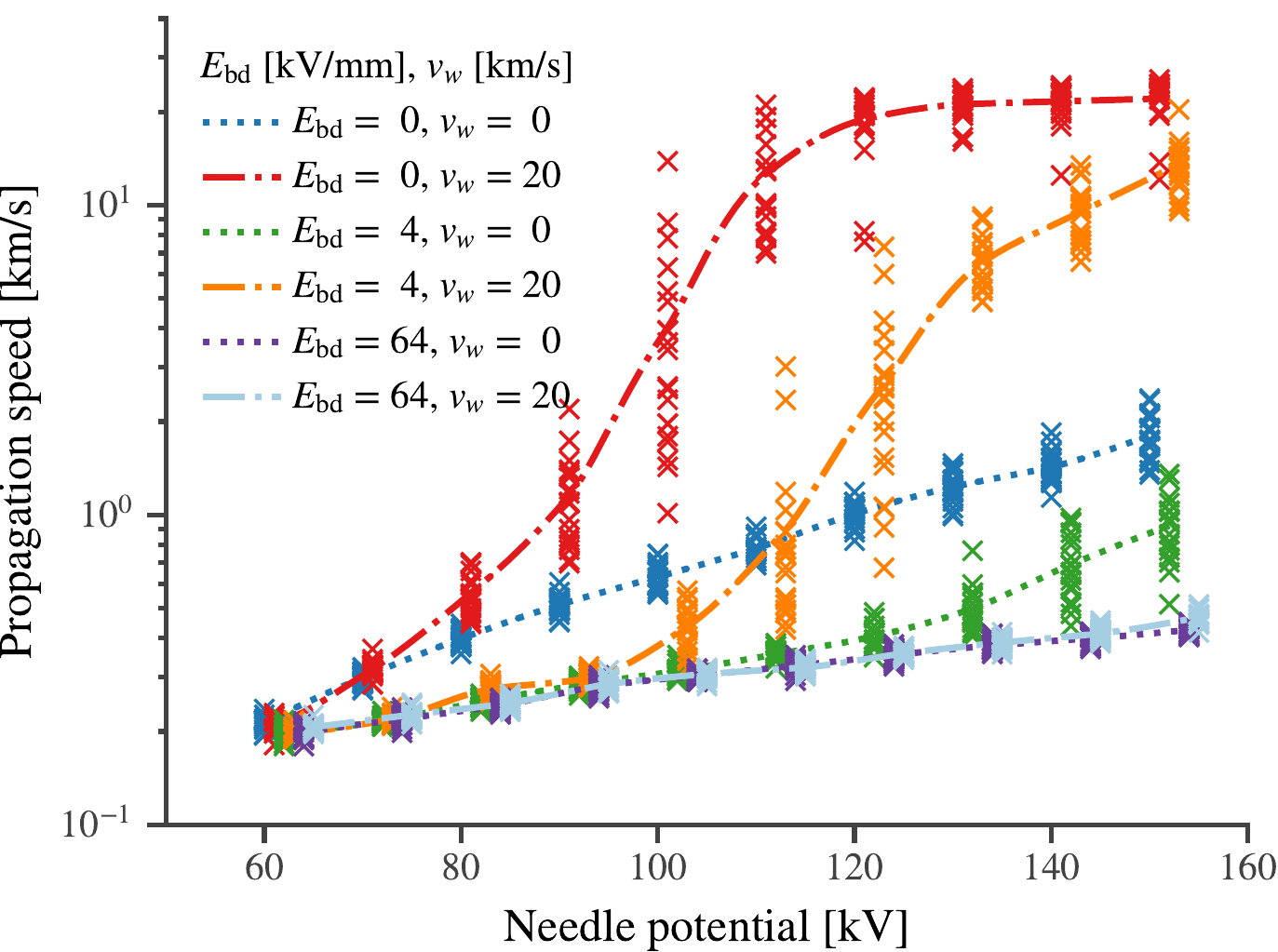}
    \caption{%
        Average propagation speed for the middle of the gap
        ($z$ between \SI{2.5}{mm} and \SI{7.5}{mm}).
        The onset of the fast mode is delayed
        when the field within the streamer channel is increased.
        Each marker is a simulation
        (20 for each voltage, 1200 in total)
        and
        the lines are interpolated to the average.
        $v_w = \SI{0}{km/s}$ implies no added speed from photoionization.
        }
    \label{fig:zapsim_streamer_lgy_v_psm_rcb(c)}
\end{figure}

For evaluating the model we investigate the influence of
the applied voltage $V_0$ (square wave),
the threshold for breakdown in the channel $E_'bd'$,
while
excluding or including photoionization.
\Cref{fig:move_photo_V120kV_d10mm}
illustrates the behavior of two different single head streamers.
Streamer 1 starts in a fast mode,
but after propagating some \si{mm}
the electric field at the streamer head
has dropped below the threshold for fast propagation $E_w$
and the streamer changes to a slower mode of propagation.
Streamer 2 starts in a slow mode,
but having no potential drop within the streamer channel,
the electric field at the streamer head increases during propagation
and the streamer changes to a fast mode for the final few \si{mm} of the gap.

Both streamer 1 and 2 in \cref{fig:move_photo_V120kV_d10mm}
are simplified cases with a single head and a constant $E_'s'$,
however,
the simulations in \cref{fig:streaks_potential}(a)
show a similar behavior,
but at higher voltages.
In the simulations with low $E_'bd'$,
resulting in a low $E_'s'$,
the streamers switch to fast mode for the final portion of the gap,
and the portion increases with increasing voltage.
According to \cref{fig:move_photo_V120kV_d10mm},
all of the streamers in \cref{fig:streaks_potential}(a)
starts above the threshold of $E_w = \SI{3.1}{GV/m}$,
however,
as the streamer propagates and more streamer heads are added,
electrostatic shielding between the heads
quickly reduces the electric field below this threshold.
Increasing $E_'bd'$ gives an on average higher $E_'s'$
and \cref{fig:streaks_potential}(a)
illustrates how this can make streamers change between fast and slow propagation.
\Cref{fig:streaks_potential}(b)
details a streamer beginning in fast mode and changing to slow propagation mode.
Propagation reduces the potential at the streamer head.
When the electric field at the tip is sufficiently reduced,
the streamer changes to a slow mode.
Re-illuminations, breakdowns within the streamer channel,
sporadically increases the potential and can push the streamer over in a fast mode,
however,
often this ``fast mode'' is brief and difficult to notice.

By considering a wider range of voltages
in \cref{fig:zapsim_streamer_lgy_v_psm_rcb(c)},
the transition from slow to fast mode
occurs at about \SI{100}{kV} for a highly conducting streamer.
Increasing $E_'bd'$ decreases the average (in time)
electric field at the streamer heads
and thus delays the onset of the fast mode to about \SI{120}{kV}.
\hl{%
An acceleration voltage of {\SI{120}{kV}} is consistent with longer gaps~%
\mbox{\cite{Lesaint2000c4xf84,Linhjell2011bjpvr2}},
while for shorter gaps ({\SI{5}{mm}}) about {\SI{60}{kV}}
has been found~\mbox{\cite{Linhjell2013cx9t}}.
}%
As mentioned in our previous work,
the propagation voltage predicted by the simulations
is somewhat high compared with experiments,
whereas the propagation speed is low for second mode streamers%
~\cite{Madshaven2018cxjf}.
The present work does not aim to improve on these limitations for slow streamers,
but rather demonstrate how changes between slow and fast
propagation can occur in different parts of the gap.
The propagation speed for slow-mode streamers
is about ten times of that predicted by \cref{fig:zapsim_streamer_lgy_v_psm_rcb(c)},
but the difference can be removed by assuming
a higher electron mobility
or
a higher seed density
~\cite{Madshaven2018cxjf}.

}


\section{Discussion}\label{sec:discussion}{

The role of photoionization during discharge
in liquids is difficult to assess.
For breakdown in gases,
ionizing radiation
can penetrate far into the medium,
providing seed electrons for avalanches.
While similar reasoning have been suggested for liquids,
we argue that,
given the higher density of the liquid
and the large cross section for ionizing radiation,
the penetration depth is short and photo\-ionization occurs locally.
\hl{%
Which radiation energies that are ionizing
and where they can cause direct ionization
are dependent on the electric field,
given the field-dependence of the IP
as well as the ionization cross-section.
Non-ionizing, low-energy
}%
radiation have longer range
and
can provide seed electrons through a two-step ionization process.
However,
ionization of impurities or additives are far more likely,
especially when the radiation from the base liquid can ionize them directly
or they have long-lived excited states.

Assuming that increasing the applied potential increases the amount of radiation,
it also increases generation of seed electrons for avalanches.
Seeds likely facilitates
both propagation speed and branching,
while electrostatic shielding between branches can regulate the propagation speed.
One hypothesis is that the change to a fast mode occurs
when one fast branch escapes the electrostatic shielding from the others.
If the radiation from such a branch can penetrate deep into the liquid,
energy is transported away from the streamer head,
while new seeds and subsequent avalanches can result in electrostatic shielding.
Both of these mechanisms can reduce the speed.
However,
we have presented a model where
a strong electric field makes photoionization more localized,
suppressing energy transport and branching.
This can explain how a streamer changes to a fast propagation mode
when the electric field is sufficiently strong.

The model is limited in the sense that we do not know
the actual value for the radiated power (or its energy distribution)
or the degree of ionization it takes for a streamer to propagate.
To assess the model we chose a value for the radiated power,
and showed that this would be sufficient to ionize the liquid at a reasonable rate.
Whether obtaining this radiated power is feasible remains unknown.

}


\section{Conclusion}\label{sec:conclusion}{

Emission and absorption of light is important for streamer propagation.
Radiation can transport energy away from the streamer as heat
or
create free electrons through ionization,
however,
ionizing radiation is rapidly absorbed
and thus
unlikely to create seed electrons at some distance from the streamer head.
Furthermore,
since increasing the electric field
reduces the ionization potential,
it also increases the ionization cross section,
making photoionization a local process.
The model based on the electron avalanche mechanism
in combination with modeling photoionization close to the steamer tip
is found to capture the feature of acceleration
of the streamer tip above a critical voltage.
The photoionization model
is missing a proper estimation of the spectral intensity of the radiation
as well as the resulting speed,
and this need to be investigated in the future.
Radiation and photoionization is often mentioned in streamer literature,
however,
the potential short reach of the ionizing radiation
is an important aspect
to consider
in understanding streamers in dielectric liquids.

}


\section*{Acknowledgment}{
This work has been supported by
The Research Council of Norway (RCN),
ABB and Statnett,
under the RCN contract 228850.
}

\bibliography{ms.bbl}

\end{document}